\documentclass[conference]{IEEEtran}
\usepackage{flushend}
\usepackage{microtype}
\usepackage{type1cm}
\usepackage{url}
\usepackage{amsmath}
\usepackage{amsthm}
\usepackage{amssymb}
\usepackage{bm}
\usepackage{psfrag}
\usepackage{graphicx}
\usepackage{xspace}
\usepackage{xcolor}
\usepackage{float}
\usepackage[footnotesize]{caption}
\usepackage{cite}
\IEEEoverridecommandlockouts

\begin{document}
\title{A Distributed MAC Protocol for Cooperation in Random Access Networks}

\author{\IEEEauthorblockN{Georg B\"ocherer\IEEEauthorrefmark{1},
Alexandre de Baynast\IEEEauthorrefmark{2}
and
Rudolf Mathar\IEEEauthorrefmark{1}
}
\IEEEauthorblockA{\IEEEauthorrefmark{1}Institute for Theoretical Information
Technology\\
RWTH Aachen University,
52056 Aachen, Germany\\ Email: \{boecherer,mathar\}@ti.rwth-aachen.de}
\IEEEauthorblockA{\IEEEauthorrefmark{2}European Microsoft Innovation Center
(EMIC) GmbH\\
Ritterstrasse 23, 52072 Aachen, Germany\\
Email: alexdeba@microsoft.com}
\thanks{This work has been supported by the UMIC Research Centre, RWTH
Aachen University.}
}

\maketitle

\begin{abstract}
WLAN is one of the most successful applications of wireless communications in
daily life because of low cost and ease of deployment. Whereas random access
schemes used in WLAN guarantee the same probability for all users to access the
channel, there still exists a significant throughput discrepancy between the
users because of their positions in the network, especially if all users are
constrained to the same energy consumption. Conversely, the farther users have
to spend much more energy than the closer users to achieve the same throughput.
In order to mitigate this discrepancy between spent energy and provided uplink
rate, this work defines a new distributed cooperative MAC protocol for two-hop
transmissions, called fairMACi. It dynamically selects the relaying
nodes while it guarantees that all users spend the same amount of energy.
Theoretical results show that fairMACi increases the minimum throughput
that can be guaranteed to all users in the network and thereby improves fairness
in terms of throughput. Monte Carlo simulations validate these results.
\end{abstract}

\section{Introduction}
The success of Wireless Local Area Network (WLAN) in day-to-day life is mainly
due to the use of a simple distributed Medium Access Control (MAC) protocol. The
\textit{carrier sense multiple access with collision avoidance} (CSMA/CA)
mechanism in \textit{distributed coordination function} (DCF) guarantees in the
long term the same opportunity of accessing the channel medium to each user in
the network independent of his channel conditions \cite{Bianchi2000}.
Additionally, multirate capabilities of IEEE 802.11 have enabled WLAN hotspots
to serve users with different channel conditions simultaneously. In the uplink,
different channel conditions however result in a strong discrepancy of the
experienced end-to-end throughput performance among the users depending on their
position in the network. Users that are facing bad channel conditions indeed
would have to spend significantly more energy than users with good channel
conditions in order to achieve the same throughput. If strict throughput
fairness between all users of the network is considered, this strategy leads to
a severe throughput degradation for the best users as shown
in~\cite{Heusse2003}. Conversely, imposing the same energy consumption for all
users significantly penalizes the throughput for the users facing bad channel
conditions.

In the recent years, cooperation in wireless networks has drawn a lot of
attention in order to mitigate throughput discrepancy between users in wireless
networks. Based on the early results presented in \cite{Cover1979}, it was shown
that cooperation among nodes for transmission has the potential to combat the
fading characteristics of wireless channels \cite{Laneman2004}. In
\cite{Sendonaris2003,Sendonaris2003a}, the authors illustrated that cooperation
between two users can be beneficial for both users.
More
recently, distributed protocols were proposed to coordinate cooperation at the
MAC layer, for instance \textit{r}DCF~\cite{Zhu2006a} and
CoopMAC~\cite{Liu2007}. Both protocols enable two-hop transmission as an
alternative to direct transmission for WLAN. These protocols also coordinate
cooperation on the PHY layer~\cite{Liu2008,Liu2008a}. The benefits of
cooperation for the whole network have been analyzed
in~\cite{Korakis2007a,Narayanan2007}. However, most of the cooperative protocols
proposed so far aim to optimize each packet flow
separately. In~\cite{Zhu2006a,Bletsas2006,Liu2007}, the authors proposed to
select the best relay for each transmission separately. However, if one node is
determined as the best relay for many nodes, its energy consumption will be very
high compared to other nodes. In~\cite{Bocherer2008} we
investigated distributed cooperative protocols for two users using DCF where
both users were constrained to achieve same throughput with same energy
consumption, i.e., full fairness. This was achieved by individual transmission
power adaption for each user. Whereas large throughput gains were observed with
this approach, the extension to scenarios with many users is unrealistic since
the resulting transmission powers vary by orders of magnitude, which is
incompatible with the typical characteristics of a power amplifier in a
transmitter.

In this paper, we assume equal transmission power for all users. We propose the
protocol fairMACi, which is designed to maximize the minimum throughput
(min-throughput) achieved by any user in the network, assuming an equal energy
constraint per user. For the uplink to a common
access point (AP), the protocol enables cooperative transmission (Two-Hop or
Decode-and-Forward) as an alternative to direct transmission. Along the lines
of~\cite{Biswas2005}, fairMACi dynamically determines the relays when
the source broadcasts its packet. Compared to DCF, fairMACi adds in
the broadcast phase a flag (1 byte) to the control overhead in each packet. This
flag contains the current SNR value between the source node and the destination.
Based on the SNR value, each node that can decode the packet decides to relay
or not the information. The estimation of other nodes' SNR and the maintenance
of a ``coopTable'' with rate information of other nodes as in \cite{Liu2007} is
not necessary in fairMACi.

The remainder of the paper is organized as follows. In Section~\ref{sec:model},
we describe the considered system setup and review the communication schemes
Direct-Link, Two-Hop and Decode-and-Forward. The corresponding MAC protocols are
defined in Section~\ref{sec:protocol}. We analyze in
Section~\ref{sec:throughput} the resulting min-throughput and discuss simulation
results.

\section{Transmission Model}
\label{sec:model}
We consider a network of $N$ randomly distributed nodes that seek to transmit
their data to a common AP. With each pair of nodes $k,l$ of the
network, we associate an achievable rate $R_{k,l}$. We denote by $R_k$ the
maximum achievable rate of the direct link from node $k$ to the AP. In our
protocol described in the next section, we assume that node~$k$ knows the
rate~$R_k$ or an estimate of it but none of the rates 
$R_{k,l},\forall k,l,k\neq l$. In comparison to CoopMAC~\cite{Liu2007}, there
is no need for each node to maintain a table referred as ``coopTable'', which
contains estimates of all rates~$R_{k,l}$. This assumption is fundamental for
the implementation point of view since it considerably reduces the amount of
information exchange between the nodes. We assume that the rates of the links
remain constant during the transmission of a few consecutive packets. We
assume continuous rate adaptation for all considered transmission schemes and
identify the achievable rate $R_{k,l}$ with the mutual information between sent
and received signal as a function of the signal-to-noise-ratio (SNR) at the
receiver $l$, i.e.,
\begin{subequations}
\begin{align}
R_{k,l} = \log(1+\mathrm{SNR}_{k,l})\quad\left[\mathrm{bits}/\mathrm{s}/\mathrm{Hz}\right].
\label{eq:Rkl}\\
R_{k} = \log(1+\mathrm{SNR}_{k})\quad\left[\mathrm{bits}/\mathrm{s}/\mathrm{Hz}\right].
\label{eq:Rk}
\end{align}
\label{eq:Rlk_and_Rk}
\end{subequations}
The SNR is defined as the ratio between the transmission power (which we
assume to be the same for all nodes) and the noise power times the
attenuation factor of the signal between $k$ and $l$ (or $k$ and the AP,
respectively). The noise is assumed to be complex symmetric additive white
Gaussian.

Since we want to guarantee equal energy consumption for all nodes, we set all
packets to the same size. We normalize it to one without loss of
generality. The amount of information that can be associated with one packet
depends on the corresponding transmission rate. The aim is to guarantee a
minimum amount of information $D$ per packet to all users in the network.
We next recall briefly the three basic transmission schemes that we consider in
this paper:~``Direct-Link'', ``Two-Hop'', and ``Decode-and-Forward.''
The three schemes are designed for two nodes $k$ and $l$ that seek to transmit
messages to the AP. They will be the cornerstones when we design our protocol
for more than two nodes in Section~\ref{sec:protocol}.

\subsubsection{Direct-Link}
Each node transmits its data directly to the AP. Since the packet size is
normalized to one, the maximum amount of information that node $k$ can transmit
to the AP within one packet is given by $R^k_{\mathrm{dir}}=R_k$. Although
$R^k_{\mathrm{dir}}$ may be larger than $D$ for some nodes $k$, the choice for $D$
is driven by the node(s) with smallest rate. Moreover, the nodes with rate
$R^k_{\mathrm{dir}}$ larger than $D$ can transmit the information $D$ in the
amount of time $t_k=D/R^k_{\mathrm{dir}} < 1$. The remaining time $1-t_k$ can be
used to support the transmission of nodes with smaller rate as in the next two
schemes.
\subsubsection{Two-Hop}
Assume node $k$ cannot achieve the target rate by directly transmitting to the
AP. Instead, node~$k$ may transmit its packet to some other closer node $l$.
Node $l$ decodes the packet, then re-encodes it and transmits it to the AP. If
node $l$ has a ``free'' amount of time $1-t_l$ at its disposal (as defined in
the previous paragraph) for forwarding the packet, the maximum rate per packet
at which node $k$ can deliver data to the AP via the relaying node $l$ is given
by
\begin{align}
R^{k,l}_{\mathrm{2hop}}=&\min\left\{R_{k,l},(1-t_l)R_l\right\}.
\label{eq:rateTwohop}
\end{align}
We denote by $H^k_\mathrm{2hop}=\{l\mid R^{k,l}_\mathrm{2hop}\geq D\}$ the
set of nodes that can effectively help $k$ to achieve the target rate $D$ via
Two-Hop.

\subsubsection{Decode-and-Forward, \cite{Cover1979}}
This scheme is similar to the two-hop scheme but it exploits the broadcast
nature of a wireless transmission. Although the AP cannot decode the
transmission of node $k$ if $R_{k} < D$, it can listen to the transmission for
``free'', record it and use it when the relaying node $l$ will forward the
message. The Decode-and-Forward scheme exploits this fact as follows. Instead of
forwarding the whole packet as in the Two-Hop scheme, the relaying node $l$ only
forwards the part of data that is missing at the AP to decode the original
transmission of node $k$. The maximum rate for this scheme is given by
\begin{align}
R^{k,l}_{\mathrm{df}}&=\min\left\{R_{k,l},R_k+(1-t_l)R_l\right\}.
\label{eq:rateDf }
\end{align}
This rate is a special case of~\cite[Prop. 2]{Host-Madsen2005}, since $k$
remains silent when $l$ is forwarding. By
\text{$H^k_\mathrm{df}=\{l\mid R^{k,l}_\mathrm{df}\geq D\}$}, we denote the set
of nodes that can help $k$ to achieve the
target rate $R$ via Decode-and-Forward. The amount of data that node $l$ has to
forward is given by $D-R_k$ and varies with $R_k$. It is strictly less than in
the Two-Hop scheme, where the amount of data to forward is always equal to $D$.

\section{MAC Protocols}
\label{sec:protocol}
\setcounter{subsubsection}{0} 
In the previous section, we reviewed the two schemes Two-Hop and
Decode-and-Forward that imply node cooperation (through the relaying node) and
saw that they could increase the target rate $D$ supported by the farther
nodes. However, we did not address the problem of \emph{coordination}, which
consists for a farther node in selecting a closer node that can help. In this
section, we introduce a new cooperative protocol, named
fairMACi which dynamically selects a ``good'' relay node based
on the current channel conditions. We shall show that this protocol increases
the target rate $D$ achievable by all nodes of the network while keeping the
energy consumption constant over the nodes.

We are interested in maximizing the minimum throughput $D$ achievable by all
nodes, which occurs when the network is in saturation (all nodes are always
backlogged). Under this assumption, the DCF of IEEE 802.11 can be modeled as a
simple CSMA scheme as shown in \cite{Bianchi2000}. In the sequel, we therefore
use this model. Additionally, we assume that the nodes are close enough to each
other such that every node can sense ongoing transmissions of any other node. We
also assume that the packet headers and the acknowledgments (ACK) are encoded at
a rate sufficiently low such that they can be decoded by any node in the
network, even when the corresponding data packet cannot be decoded.
Finally, we neglect the collisions with ACKs, i.e., all ACKs from the AP will be
detected and decoded correctly by all nodes in the network. Assuming the
mechanisms at the physical layer described in Section~\ref{sec:model}, we detail
the three transmission schemes from the MAC layer perspective. CSMA with
Direct-Link is used as reference for our new protocol fairMAC{i}, which
enables Two-Hop or Decode-and-Forward.
\subsubsection{Direct-Link}
When node $k$ seeks to transmit a packet, it competes for the medium according
to CSMA:~if $k$ senses the channel idle, it initiates a transmission with
probability $\tau$. If no other node is transmitting meanwhile, the AP can
decode the packet and sends an ACK in return. Otherwise, a collision occurs; no
ACK is sent by the AP; Node $k$ declares its packet lost and will try to
transmit again the same packet later.

\subsubsection{Two-Hop~fairMAC{i}}
The transmission of a packet via Two-Hop can be split into two phases, the
broadcast phase and the relay phase. The relay phase happens only if the AP
could not decode the packet at the end of the broadcast phase. Assume that node
$k$ accesses the channel. Node $k$ starts to broadcast a packet $p_k$ at
target rate $D$.
\begin{itemize}
\item[1a)] In case of collision, no node can decode $p_{k}$ and $k$ competes again for the channel.
\item[1b)] If no collision occurs, all nodes within its transmission range successfully decode $p_{k}$ and record it.
    \begin{itemize}
    \item[1b1)] If node $k$ is close to the AP ($R_k\geq D$), the AP
successfully decodes $p_k$ and sends an ACK back to node $k$.
            All nodes receive the ACK and discard the recorded signal.
    \item[1b2)] If node $k$ is far from the AP ($R_k < D$), the AP cannot decode
$p_k$, but it stores $p_k$ and sends an ACK indicating that there was no 
collision. The transmission of $p_k$ enters the relay phase, which is
described next.
    \end{itemize}
\end{itemize}
After the broadcast phase, all nodes in $H_\mathrm{2hop}^k$ decode $p_k$
successfully. Each node $l\in H_\mathrm{2hop}^k$ forms a joint packet
consisting of $p_k$ (amount of data equal to $D$) and own data (amount of
data equal to $R_l-D\geq D$) and puts it in its packet queue. The header
of the joint packet contains in addition the MAC address of $k$ and the packet
number corresponding to $p_k$. The relay phase starts. All nodes compete for the
channel. Assume that node $l\in H_\mathrm{2hop}^k$ obtains the channel access
and the joint packet containing $p_k$ is first in its packet queue.
The relay phase begins when node $l$ starts to transmits this joint packet to
the AP.
\begin{itemize}
\item[2a)] In case of collision, all nodes in $H_\mathrm{2hop}^k$ keep the joint
packets with data $p_k$ in their queues and continue to compete for the channel.
\item[2b)] Otherwise, the AP sends an ACK to $l$ and $k$ for the successful
reception of the joint packet. Under our assumptions, all nodes in the network can decode the ACK.
All nodes in $H_\mathrm{2hop}^k$ but $l$ remove $p_k$ from the corresponding joint packet in their packet queues; node $l$
removes the whole joint packet from its queue, and node $k$ removes $p_{k}$ from its queue.
\end{itemize}

\subsubsection{Decode-and-Forward~fairMAC{i}}
The MAC protocol for Decode-and-Forward is similar to the one for Two-Hop.
However, nodes in $H^k_\mathrm{df}$ relay only the information of $p_k$ that is
missing at the AP. This operation requires the knowledge of the information
received by the AP during the broadcast phase. Under our assumptions, this
information depends only on the rate $R_{k}$. This approach requires to add
$R_k$ (or a quantized version of it) to the header of each packet $p_k$ sent by
node $k$ as we suggested in the introduction. (Alternatively, node $l$ can try
to estimate $R_k$ by its own as proposed in~\cite{Liu2007}.) With this
modification, the protocol for Decode-and-Forward follows the lines of the
Two-Hop protocol.

For Direct-Link, if node $k$ is very far from the AP, that is, if
$R^k_\mathrm{dir}<D$, node $k$ is not physically supported by the network. In
this case, we assume that node $k$ remains silent forever.

In fairMACi, node $k$ broadcasts its packet at rate $D$
hoping that the AP can decode it or at least that some other node(s) in the network can
relay the packet. If there is no such node in the network, a successful
decoding of packet $p_k$ will never be acknowledged by the AP. To prevent node
$k$ from flooding the network with additional transmissions, we impose that it
broadcasts only up to $Q$ successive packets before receiving an ACK for the
first one.
\begin{figure*}
\footnotesize
\centering
\begin{minipage}{0.26\textwidth}
\includegraphics[width=\textwidth]{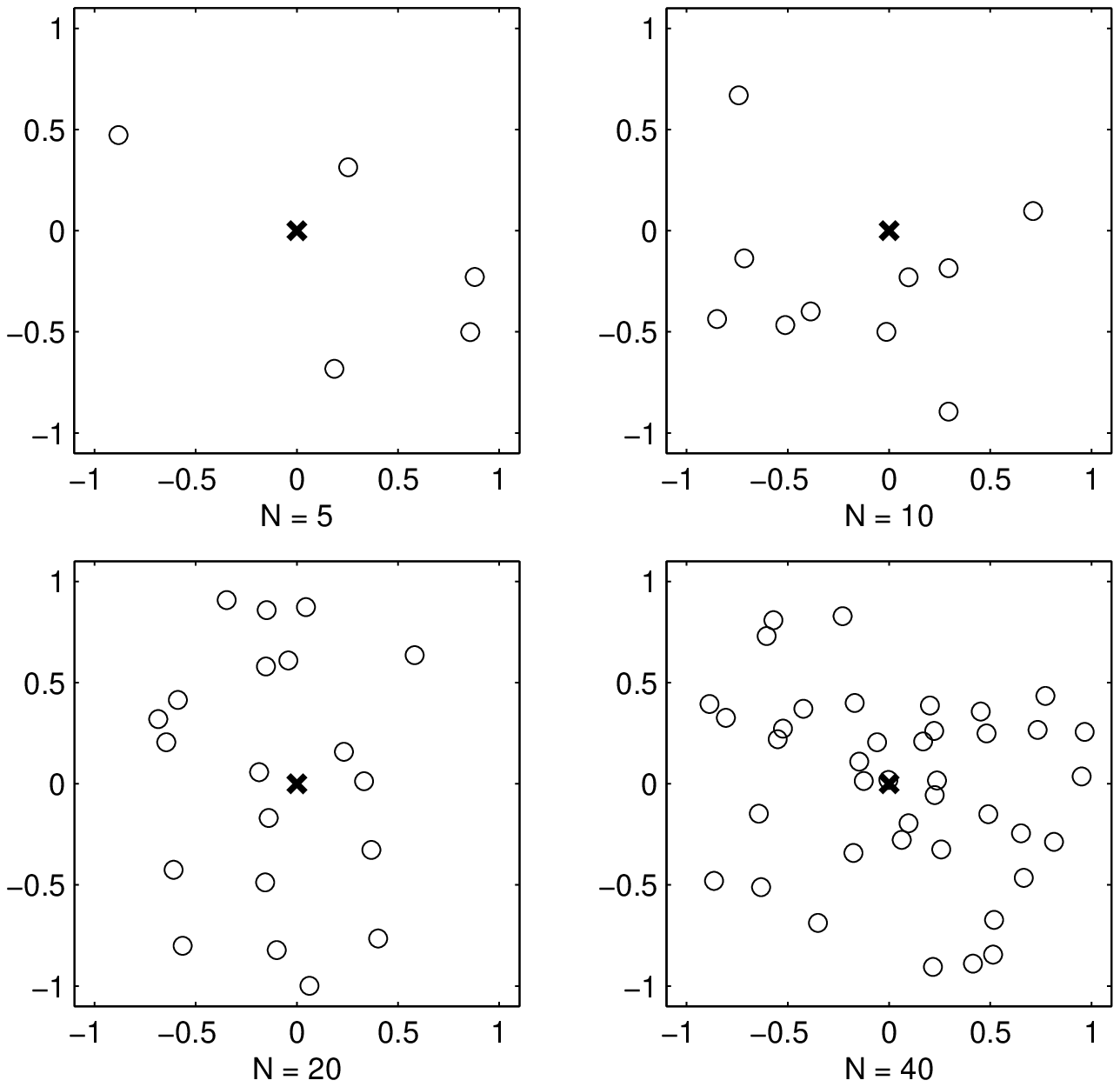}
\caption{Random topologies with 5, 10, 20, and 40 nodes.}
\label{fig:randomTopology}
\end{minipage}
\hfill
\begin{minipage}{0.34\textwidth}
\includegraphics[width=\textwidth]{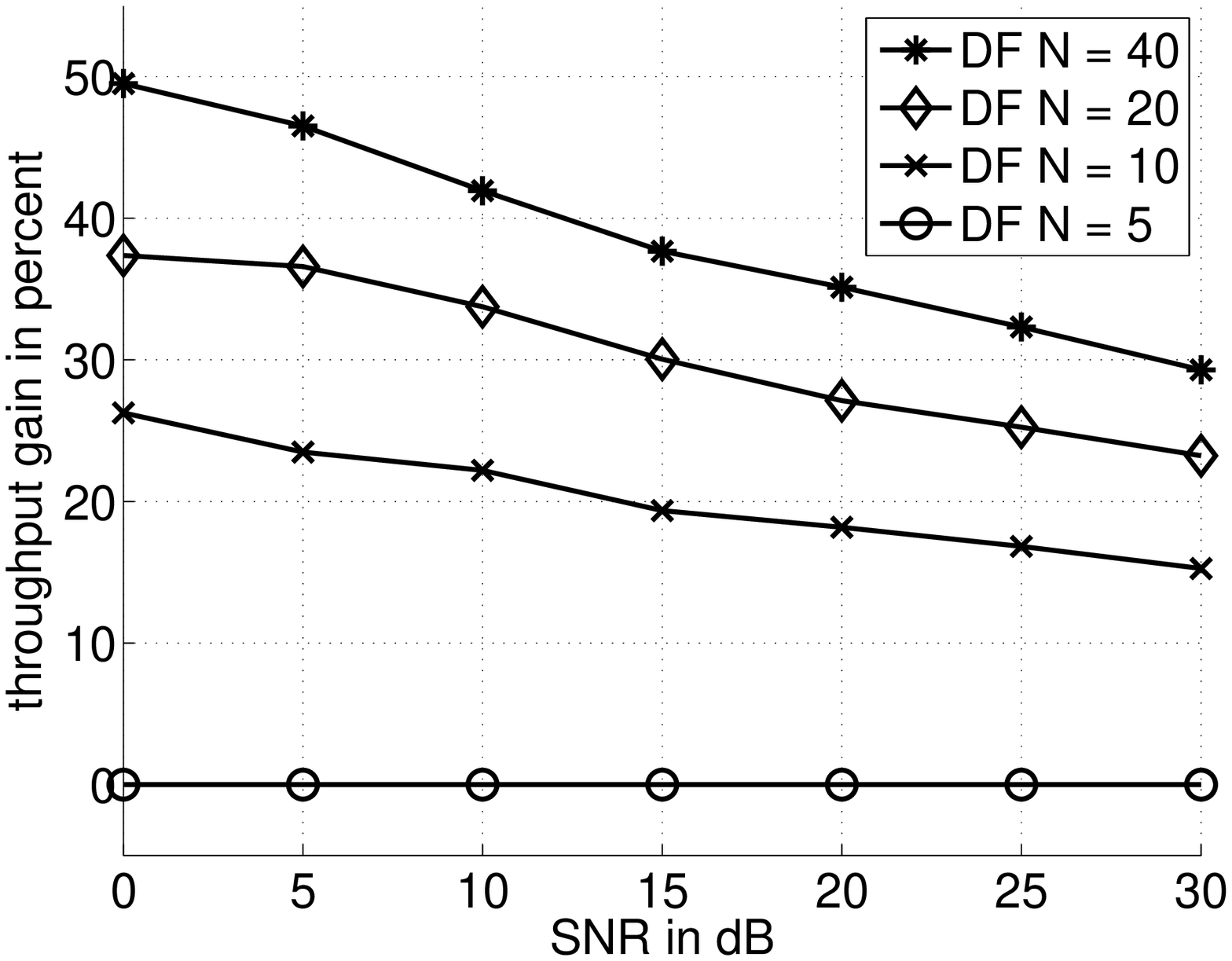}
\caption{Effective throughput gain over Direct-Link of Decode-Forward.}
\label{fig:throughputDF}
\end{minipage}
\hfill
\begin{minipage}{0.34\textwidth}
\includegraphics[width=\textwidth]{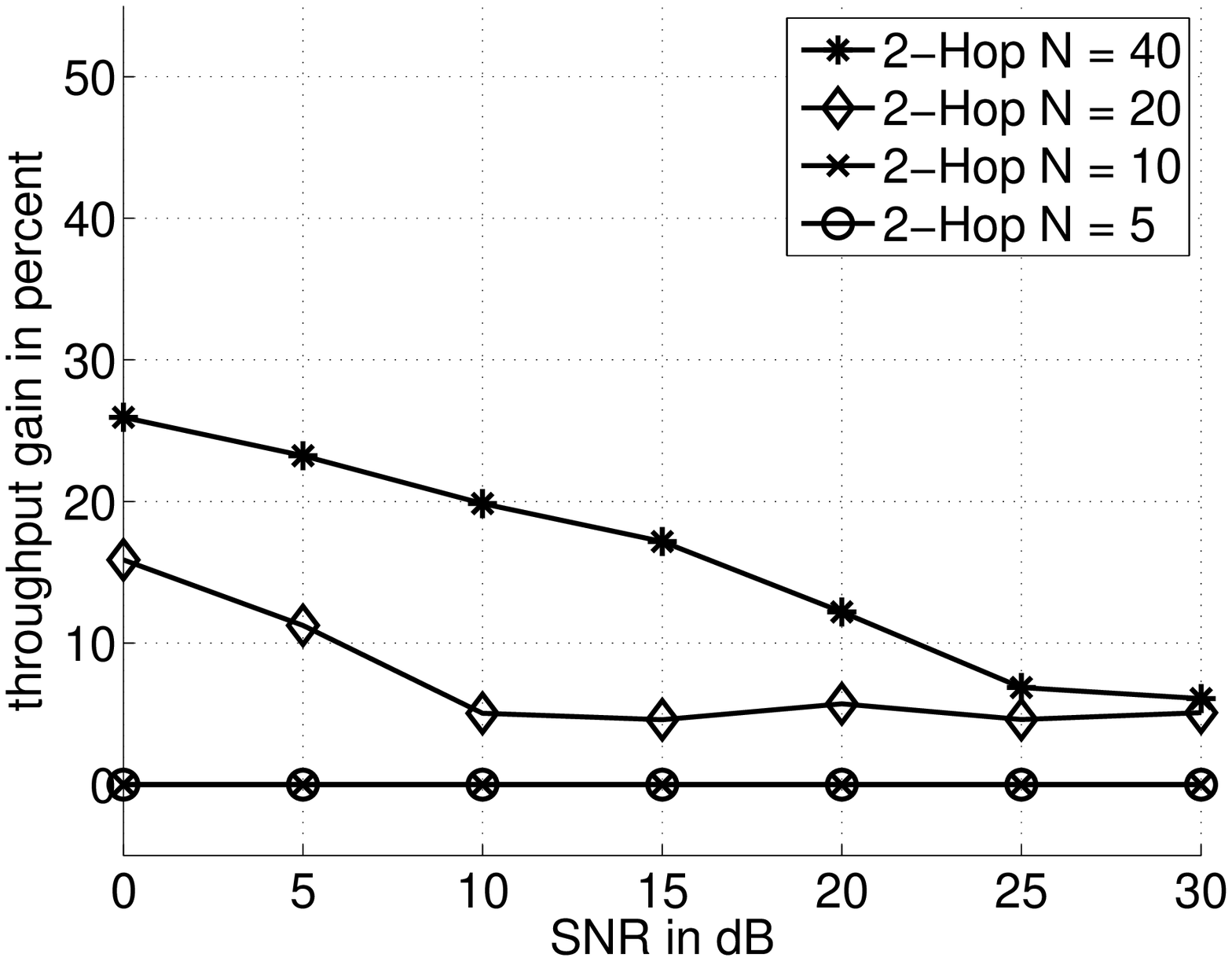}
\caption{Effective throughput gain over Direct-Link of Two-Hop.}
\label{fig:throughputTwohop}
\end{minipage}
\end{figure*}

\section{Throughput Analysis}\label{sec:throughput}
We evaluate the three protocols introduced in the previous section with respect
to fairness: under the constraint that all users spend the same amount of
energy on the long term, we measure the unfairness resulting from variations in
the data throughput provided to each user by the \emph{min-throughput}, where
the minimum is taken over all users in the network. High min-throughput
indicates a low variance of throughput over the users, which corresponds to an
increased degree of fairness. In our analysis, we assume large packets such that
the size of ACKs and packet headers is negligible.

Assume that all nodes operate in saturation mode, i.e., they are backlogged and
we do not need to consider packet arrival processes in our analysis. Also,
assume that there is no degradation on the MAC layer, that is, on the long term
all nodes have the same number of channel accesses and consequently transmit the
same number of packets to the AP. This holds for Direct-Link, Two-Hop, and
Decode-and-Forward, since forwarding is performed by forming joint packets of
fixed size one: there is no difference in terms of competition for the channel
between a standard packet and a joint packet. In addition, the same number of
transmitted packets, the common transmission power, and the uniform packet size
of one guarantee that each node spends the same amount of energy.

The effective min-throughput of the nodes can be calculated along the lines
of~\cite{Bianchi2000}. Assume that $N$ nodes compete for the channel. We slot
the time into time slots of length $\sigma$. If the channel is sensed idle
during the current time slot, every node transmits in the next time slot
independently with probability $\tau$. After an idle time slot, the
probabilities of successful transmission, idle state, and collision are given
respectively by
\begin{align}
p_s=N(1-\tau)^{N-1}\tau\mbox{; } p_i=(1-\tau)^N\mbox{; }p_c=1-(p_s+p_i).
\end{align}
A successful transmission or a collision take the amount of time one, and
both are always followed by an idle time slot. The average effective
min-throughput $S(D)$ of each node is thus given by
\begin{align}
S(D)= p_s D / \left\{ N\left[ (1-p_i)(1+\sigma)+p_i\sigma\right]
\right\}.\label{eq:throughputBound}
\end{align}
If a target rate $D$ is physically supported by a network,
\eqref{eq:throughputBound} gives an upper bound for the maximum effective
min-throughput of the network. However, even if the considered target rate is
physically supported, depending on the system
configuration \eqref{eq:throughputBound} may not be achievable because
of degradation on the MAC layer, as we discuss in subsection
\ref{subsec:deg1} and \ref{subsec:deg2}.

\section{Numerical Results}
\begin{figure*}
\footnotesize
\centering
\begin{minipage}{0.32\textwidth}
\psfrag{n1}{$n_1$}
\psfrag{n2}{$n_2$}
\psfrag{n3}{$n_3$}
\psfrag{n4}{$n_4$}
\psfrag{n5}{$n_5$}
\psfrag{ap}{Access Point}
\includegraphics[width=\textwidth]{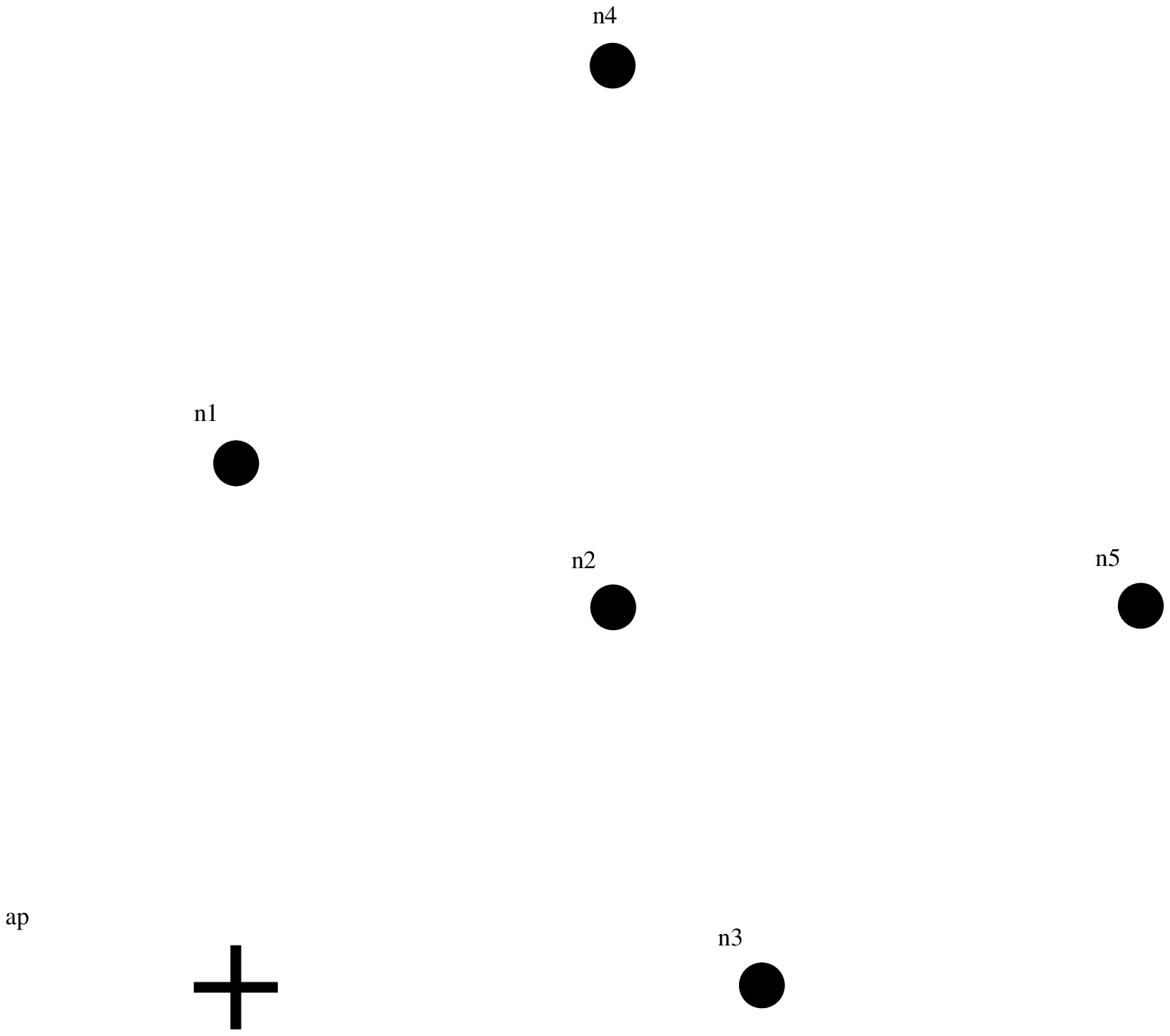}
\caption{A network with one AP and five nodes. The nodes in
$H=\{n_1,n_2,n_3\}$ can potentially help the nodes in $C=\{n_4,n_5\}$ to
deliver data to the AP.}
\label{fig:network}
\end{minipage}
\hfill
\begin{minipage}{0.32\textwidth}
\psfrag{POWER}{Power}
\psfrag{TIME}{Time}
\psfrag{memory size Q}{\footnotesize{parameter Q}}
\includegraphics[width=\textwidth]{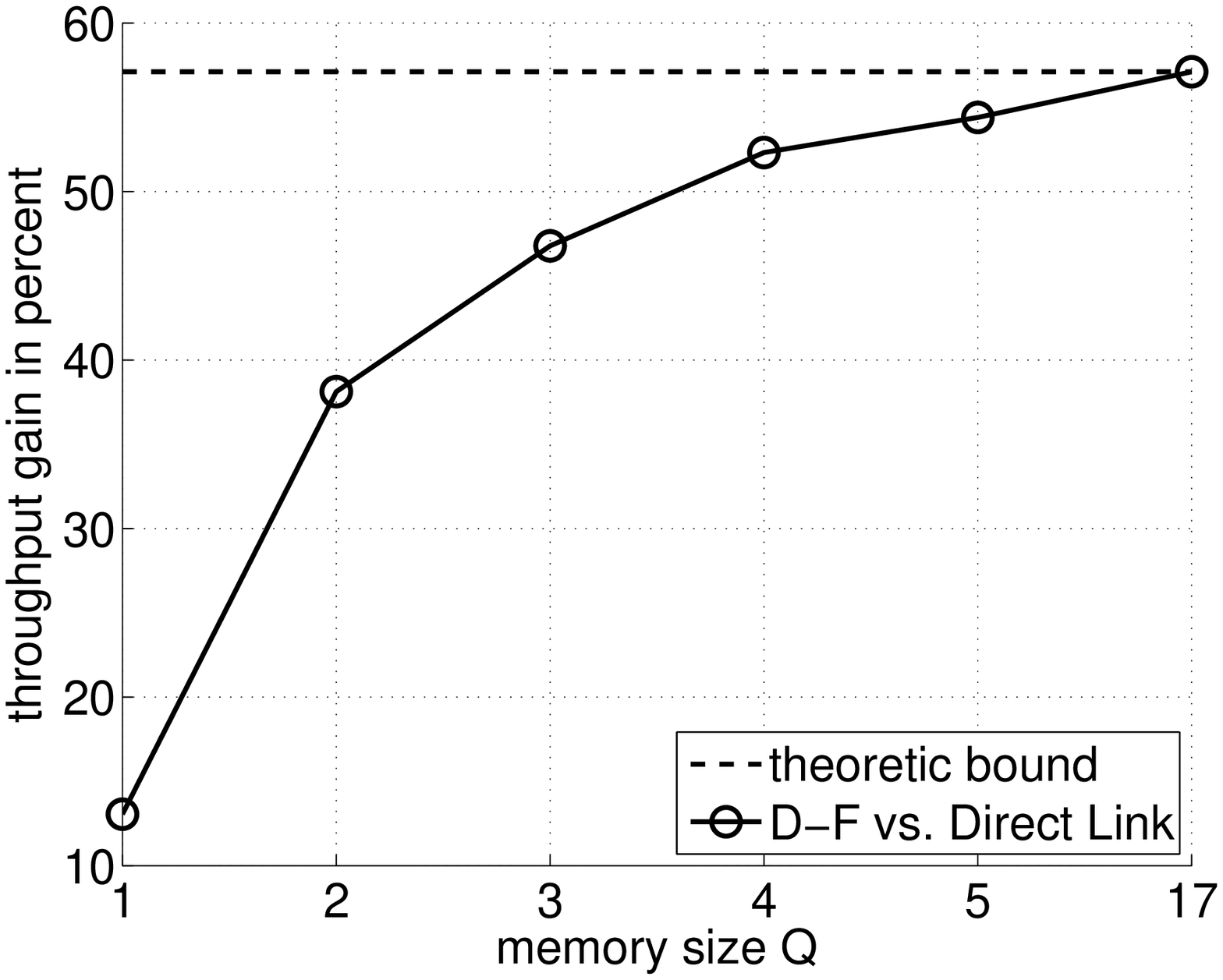}
\caption{Effective throughput gain vs. $Q$ for the network in (a). Note that
$|H|>|C|$ and that the bound \eqref{eq:throughputBound} is reached.}
\label{fig:throughputVsQ}
\end{minipage}
\hfill
\begin{minipage}{0.33\textwidth}
\psfrag{POWER}{Power}
\psfrag{TIME}{Time}
\includegraphics[width=\textwidth]{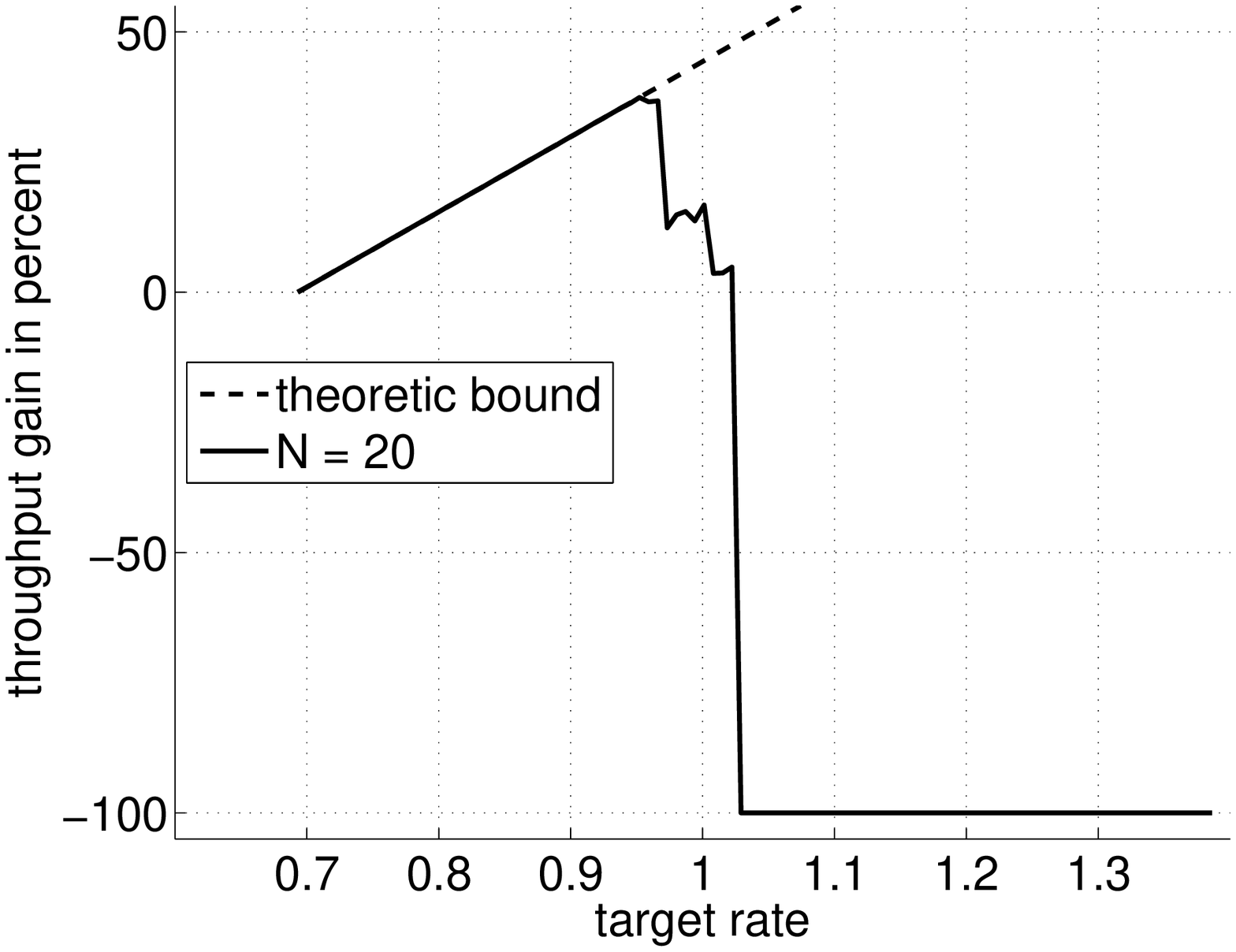}
\caption{Effective throughput gain vs. target rate for the topology with
$N=20$ in Fig.~\ref{fig:randomTopology}. Notice the regions ``no MAC
degradation'', ``MAC degradation'', and ``physically unsupported''.}
\label{fig:throughputVsRate}
\end{minipage}
\end{figure*}
We apply our protocol to the 4 random topologies in
Fig.~\ref{fig:randomTopology} with 5, 10, 20, and 40 nodes. The node positions
are uniformly distributed over the unit circle and normalized such that the node
farthest away from the AP is at distance one. We assume free-space pathloss,
i.e., the signal power is attenuated with the source-destination distance to the
power of $\gamma$ \cite{Tse2005}. We set $\gamma=2$. The transmit power is
specified in SNR at the transmitting nodes and set to the same value for all
nodes. The maximum number of unacknowledged packets is set to \text{$Q=100$}.
For all
setups, 5 millions of packets are transmitted to the AP. The transmission
probability is set to $\tau=0.001$ and the normalized time slot length is set to
$\sigma=0.002$. We maximize the effective min-throughput gains 
compared to Direct-Link
over the target rate $D$ for Decode-and-Forward and for Two-Hop. The results are
shown in Fig.~\ref{fig:throughputDF} and in Fig.~\ref{fig:throughputTwohop},
respectively, as a function of the SNR. The effective min-throughput gains
increase for both Two-Hop and Decode-and-Forward with the number of nodes in the
network. This is intuitive since more nodes can be found in ``good'' positions
for relaying in denser networks. The min-throughput gains for Decode-and-Forward
are higher than for Two-Hop, which confirms the results from
\cite{Bocherer2008}. By investigating the dependency of our results on the
parameters $Q$ (maximum number of unacknowledged packets) and $D$ (target rate),
we identified two kinds of throughput degradation on the MAC layer.

\subsection{First Kind of MAC Degradation}\label{subsec:deg1} 
To prevent the nodes from flooding the network with retransmissions in
fairMACi, we limited the number of unacknowledged packets to $Q$.
However, this can degrade the throughput on the MAC layer, since the random
access in CSMA can lead to an unbounded number of unacknowledged packets. This
degradation can be diminished by setting $Q$ to a finite but large enough value.
In Fig.~\ref{fig:throughputVsQ}, the effective min-throughput gain of
Decode-and-Forward over Direct-Link is displayed as a function of $Q$. In the
considered example, bound
\eqref{eq:throughputBound} is already reached for $Q=17$. Note that for a given
value of $Q$ and a network of $N$ nodes, the maximum amount of additional memory
needed by the relaying nodes is upper bounded by $P\leq QN$.

\subsection{Second Kind of MAC Degradation}\label{subsec:deg2}
The second kind of degradation occurs when the number of relaying nodes $|H|$ is
small compared to the number of the other nodes $|C|$. The nodes in $C$ are
waiting for $Q$ unacknowledged packets most of the time and are therefore unable
to transmit new packets. This effect is illustrated in
Fig.~\ref{fig:throughputVsRate}. For a random topology with 20 nodes, the target
rate $D$ is gradually increased. Consequently, the number of nodes in $H$
decreases and the number of nodes in $C$ increases. As long as there are enough
helping nodes, the effective throughput gain follows the theoretical
bound~\eqref{eq:throughputBound}. After reaching a certain rate, the effective
throughput gain rapidly decreases:~although the target rate is still physically
supported by the network, at least one relay node starves, since the number of
nodes that need help exceeds the number of relaying nodes. As the target rate
farther increases, the effective throughput gain drops down to -100 \%, since
there is at least one node in the network that cannot achieve the target rate by
any scheme. The target rate is no longer physically supported by the network.
The oscillating behavior of the curve is not random but depends on the topology.
A future challenge is to determine the optimum operation point of
fairMACi in a running network by choosing the target rate parameter $D$
properly.

\section{Conclusions}
In this paper, we have proposed a new distributed cooperative protocol
fairMACi for WLAN uplink transmissions that improves the
min-throughput compared to the basic DCF under constant average energy per user.
Our protocol supports Two-Hop and Decode-and-Forward transmissions. Since the
maximization problem in terms of min-throughput guaranteed to any user in the
network is equivalent to lower the variance of the individual throughput over
the users, our protocol increases fairness in terms of throughput. For a random
topology with $40$ nodes, Decode-and-Forward provides min-throughput gains over
Direct-Link of up to 50\% and Two-Hop provides min-throughput gains of more than
25\% for a large range of SNR. A possible extension of this work consists in
finding a distributed target rate adaptation protocol that maximizes the
min-throughput of fairMACi. Also, practical coding schemes
should be addressed in order to determine if the theoretical gains observed in
the present work are achievable in real networks.

\bibliographystyle{IEEEtran}

\bibliography{IEEEabrv,confs-jrnls,Literatur}

\end{document}